%% file: mainRita2.tex
\documentclass[twocolumn, 5p]{article} 
\usepackage[utf8]{inputenc}

\usepackage{lineno,hyperref}
\modulolinenumbers[5]

\usepackage{url}
\usepackage[a4paper,margin=1in]{geometry}
\usepackage{enumitem}
\usepackage{acro}
\usepackage{graphicx}
\usepackage{epsfig}
\usepackage{natbib}
\usepackage{multirow}
\usepackage{lscape}
\usepackage{longtable}
\usepackage{tikz}
\usepackage{array}
\usepackage{color,soul}
\usepackage{authblk}

\DeclareAcronym{RGC}{short=RGC, long=retinal ganglion cell}
\DeclareAcronym{ONH}{short=ONH, long=optic nerve head}
\DeclareAcronym{RNFL}{short=RNFL, long=retinal nerve fiber layer}
\DeclareAcronym{BMO-MRW}{short=BMO-MRW, long=Bruch's membrane minimum-rim-width}
\DeclareAcronym{LC}{short=LC, long=lamina cribrosa}
\DeclareAcronym{IIH}{short=IIH, long=idiopathic intracranial hypertension}
\DeclareAcronym{ON}{short=ON, long=optic neuritis}
\DeclareAcronym{MS}{short=MS, long=multiple sclerosis}
\DeclareAcronym{NMOSD}{short=NMOSD, long=neuromyelitis optica spectrum disorders}
\DeclareAcronym{OCT}{short=OCT, long=optical coherence tomography}
\DeclareAcronym{SD-OCT}{short=SD-OCT, long=spectral-domain optical coherence tomography}
\DeclareAcronym{EDI}{short=EDI, long=enhanced depth imaging}
\DeclareAcronym{SS-OCT}{short=SS-OCT, long=swept-source optical coherence tomography}
\DeclareAcronym{IOP}{short=IOP, long=intraocular pressure}
\DeclareAcronym{CSFP}{short=CSFP, long=cerebrospinal fluid pressure}
\DeclareAcronym{TLPG}{short=TLPG, long=translaminar pressure gradient}
\DeclareAcronym{BM}{short=BM, long=Bruch's membrane}
\DeclareAcronym{BMO}{short=BMO, long= Bruch's membrane opening}
\DeclareAcronym{RPE}{short=RPE, long=retinal pigment epithelium}
\DeclareAcronym{SVM}{short=SVM, long=support vector machine}
\DeclareAcronym{GMM}{short=GMM, long=Gaussian mixture model}
\DeclareAcronym{GAN}{short=GAN, long=generative adversarial network}
\DeclareAcronym{CNN}{short=CNN, long=convolutional neural network}
\DeclareAcronym{DRUNET}{short=DRUNET, long=Dilated-Residual U-Net}
\DeclareAcronym{MRF}{short=MRF, long=Markov Random Field}
\DeclareAcronym{TBMR}{short=TBMR, long=three bench mark reference}
\DeclareAcronym{CDR}{short=CDR, long=cup to disc ratio}
\DeclareAcronym{ILM}{short=ILM, long=inner limiting membrane}
\DeclareAcronym{AUC}{short=AUC, long=area under the curve}
\DeclareAcronym{ANN}{short=ANN, long=artificial neural network}
\DeclareAcronym{PCA}{short=PCA, long=principal component analysis}
\DeclareAcronym{NN}{short=NN, long=nearest neighbor}
\DeclareAcronym{POAG}{short=POAG, long=primary open angle glaucoma}
\DeclareAcronym{GODA}{short=GODA, long=glaucomatous \ac{ONH} appearence}
\DeclareAcronym{RMSE}{short=RMSE, long=root mean square error}
\DeclareAcronym{OD}{short=OD, long=optic disc}

\begin{document}
\date{}
\title{Automatic Segmentation of the Optic Nerve Head Region in Optical Coherence Tomography: A Methodological Review}

\author[1,2]{Rita Marques}
\author[2,*]{Danilo Andrade De Jesus}
\author[3,5,6]{João Barbosa Breda}
\author[3,4]{Jan Van Eijgen}
\author[3,4]{Ingeborg Stalmans}
\author[2]{Theo van Walsum}
\author[2]{Stefan Klein}
\author[1]{Pedro G. Vaz}
\author[2]{Luisa Sánchez Brea}

\affil[1]{Laboratory for Instrumentation, Biomedical Engineering and Radiation Physics (LIBPhys-UC), Department of Physics, University of Coimbra, Coimbra, Portugal}
\affil[2]{Biomedical Imaging Group Rotterdam, Department of Radiology \& Nuclear Medicine, Erasmus MC, Rotterdam, Netherlands}
\affil[3]{Research Group Ophthalmology, Department of Neurosciences, KU Leuven, Leuven, Belgium}
\affil[4]{Department of Ophthalmology, University Hospitals UZ Leuven , Leuven, Belgium}
\affil[5]{Cardiovascular R\&D Center, Faculty of Medicine of the University of Porto, Porto, Portugal}
\affil[6]{Ophthalmology Department, Centro Hospitalar e Universitário São João, Porto, Portugal}
\affil[*]{Corresponding author: d.andradedejesus@erasmusmc.nl}

 \twocolumn[ \begin{@twocolumnfalse}
  \maketitle
    \begin{abstract}
    The \ac{ONH} represents the intraocular section of the optic nerve, which is prone to damage by \ac{IOP}.  The advent of \ac{OCT} has enabled the evaluation of novel \ac{ONH} parameters, namely the depth and curvature of the \ac{LC}. Together with the \ac{BMO-MRW}, these seem to be promising \ac{ONH} parameters for diagnosis and monitoring of retinal diseases such as glaucoma. Nonetheless, these \ac{OCT} derived biomarkers are mostly extracted through manual segmentation, which is time-consuming and prone to bias, thus limiting their usability in clinical practice. The automatic segmentation of \ac{ONH} in \ac{OCT} scans could further improve the current clinical management of glaucoma and other diseases.
    This review summarizes the current state-of-the-art in automatic segmentation of the \ac{ONH} in \ac{OCT}. PubMed and Scopus were used to perform a systematic review. Additional works from other databases (IEEE, Google Scholar and ARVO IOVS) were also included, resulting in a total of 27 reviewed studies. 
    For each algorithm, the methods, the size and type of dataset used for validation, and the respective results were carefully analyzed. The results show that deep learning-based algorithms provide the highest accuracy, sensitivity and specificity for segmenting the different structures of the \ac{ONH} including the \ac{LC}. However, a lack of consensus regarding the definition of segmented regions, extracted parameters and validation approaches has been observed, highlighting the importance and need of standardized methodologies for \ac{ONH} segmentation.
    \end{abstract}
\end{@twocolumnfalse}]

\section{Introduction}

\ac{OCT} is an imaging technique that enables noninvasive cross-sectional imaging of tissue using low-coherence interferometry. Among its many applications, one of the most common is the analysis of the human retina where, given its high resolution and three-dimensional nature, \ac{OCT} can assist in the diagnosis and prognosis of several diseases.

One example of such diseases is glaucoma, which is the main cause of irreversible blindness worldwide, and which has elevated \ac{IOP} as primary risk factor for its development \citep{Sigal2005}. It starts to manifest through damage to the \ac{RGC} axons as they exit the eye at the \ac{ONH} \citep{Tan2018}, and is associated with complex 3D structural modifications in the \ac{ONH}, such as thinning of the \ac{RNFL}, changes in the \ac{BMO-MRW} and in the \ac{LC} depth, thickness and curvature \citep{Tan2018, bekkers2020microvascular, Lee2017, de2020octa, jesus2019quantitative}.  Evidence suggests that the \ac{ONH} surface depression occurs before the \ac{RNFL} thinning \citep{Xu2014}, making it more relevant for early diagnosis.

Even though \ac{ONH} structural changes have been mostly studied in a context of glaucoma diagnosis, they are also representative of non-ophthalmic diseases such as \ac{IIH}, \ac{ON}, \ac{MS} or \ac{NMOSD} \citep{Yadav2018}, Alzheimer \citep{CabreraDeBuc2019, lemmens2020combination}, and Parkinson's disease \citep{Eraslan2016}.

The \ac{LC} is a mesh-like structure that fills the posterior scleral foramen where unmyelinated \ac{RGC} axons pass through before exiting the eye \citep{Tan2018}. Up until recently, it has been difficult to study this region of the \ac{ONH}, given its deep location. The \ac{OCT} signal is highly attenuated when reaching deeper structures, and the shadow of the blood vessels, which merge at the \ac{ONH}, can limit the correct identification of the \ac{LC} and other \ac{ONH} structures \citep{Yadav2018}. However, it is now possible to overcome some of these problems thanks to advances in imaging technologies such as \ac{EDI} \citep{Park2012}, \ac{SS-OCT} \citep{Takusagawa2019} and adaptive compensation \citep{Mari2013}.

Given the diagnostic potential of biomarkers extracted from the \ac{ONH}, and particularly from the \ac{LC} \citep{Paulo2021}, an accurate segmentation of this region is becoming increasingly important for improving clinical diagnosis and follow-up, and also for contributing to a better understanding of several ophthalmic and non-ophthalmic diseases.

The need for an automatic segmentation arises from the fact that manual segmentation is time consuming, prone to bias, and unsuitable for a clinical environment \citep{Lang2013} since it requires extensive training and expertise. Even if some commercial \ac{OCT} devices already have an in-built proprietary segmentation software, they can segment some, but not all, \ac{ONH} tissues, and they still require frequent manual corrections \citep{Devalla2020}.

The democratization and emergence of \ac{OCT} as the clinical gold-standard for in vivo structural ophthalmic examinations \citep{Fujimoto2016} has encouraged the entry of new manufacturers to the market as well. It will soon become practically infeasible to perform manual segmentations for all \ac{OCT} brands, device models, generations, and applications \citep{Devalla2020}. From this arises the urgent need for device-independent automatic segmentation algorithms.

This review summarizes the current state-of-the-art in automatic segmentation of the \ac{ONH} in \ac{OCT}.

\section{Methods}

A literature search was conducted in MEDLINE (Pubmed) and Scopus bibliographic databases on the 24\textsuperscript{th} of November 2020. The search query (PubMed) was: (imag* AND processing OR segmentation) AND (optic AND (disc OR disk) OR lamina) AND (optical AND coherence AND tomography) NOT (fundus OR angiography).
Additional works which were not found by Pubmed and Scopus, but were cited in the bibliography, and could be found by Google Scholar, IEEE or ARVO bibliographic databases, were also added.
Only articles published in English with a detailed description of the method used were considered, and no publication date restriction was added. The exclusion criteria were: (i) no description of a novel segmentation algorithm; (ii) no segmentation of the \ac{ONH}; (iii) only used \emph{en-face} images; (iv) review article; (v) case report; (vi) comparative study; (vii) clinical trial; or (viii) not an article (abstracts, book chapters, editorials, and notes). 

\section{Results}
\begin{figure}[t]
\centering
\includegraphics[width=\linewidth, height=7cm]{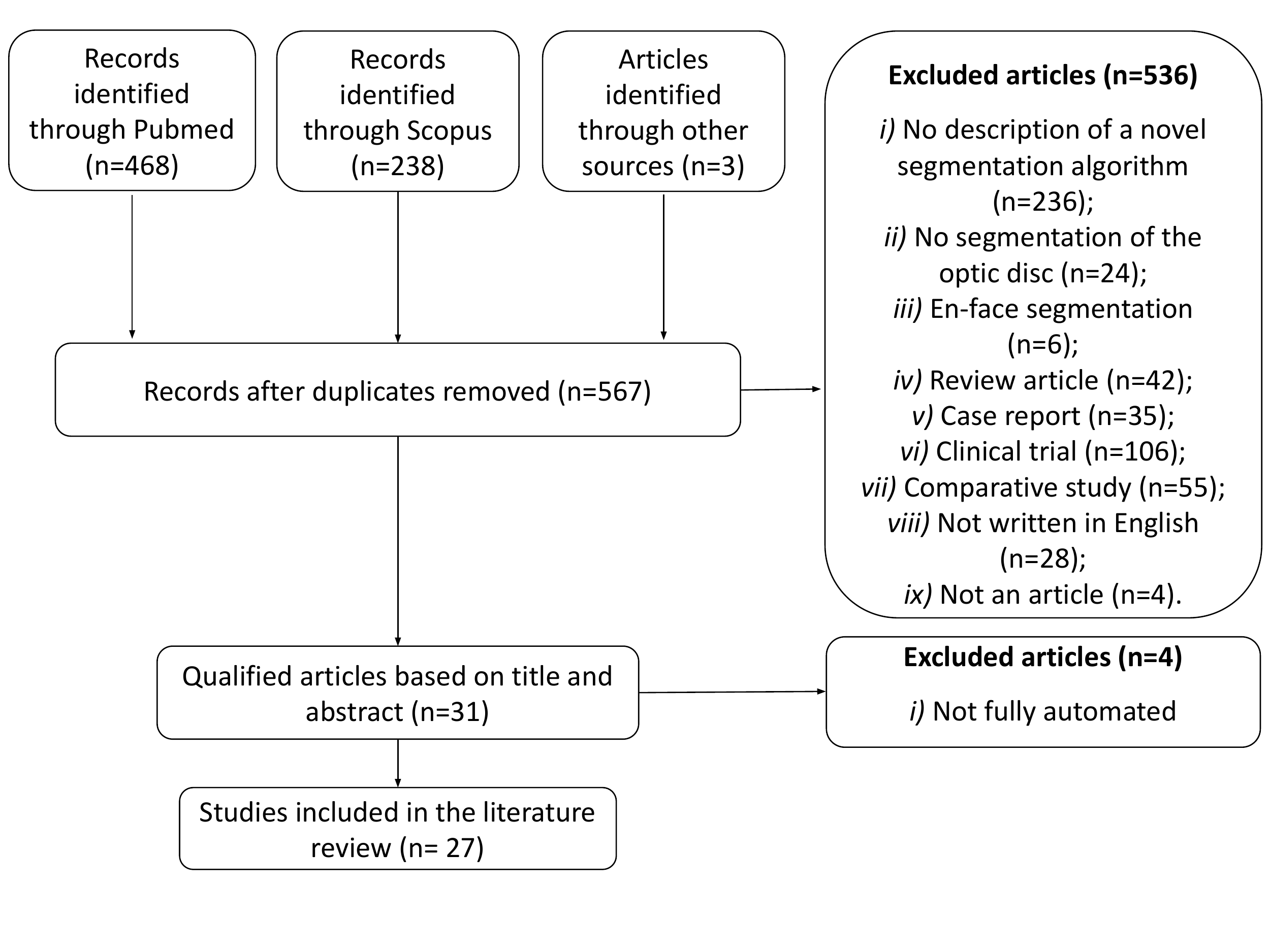}
\caption{Flowchart of the collected data.}
\label{fig:flowchart}
\end{figure}

The systematic search led to a total of 565 references after removing the duplicates, which were narrowed down to 31 after title/abstract screening, and finally to 27 after a full-text screening (Figure \ref{fig:flowchart}). 
The 27 included studies provided the description of a fully automatic segmentation of \ac{ONH} centered \ac{OCT} B-scans and/or volumes, and described how its performance was evaluated. 

\begin{figure}[t]
\centering
\includegraphics[width=\linewidth, height=8cm]{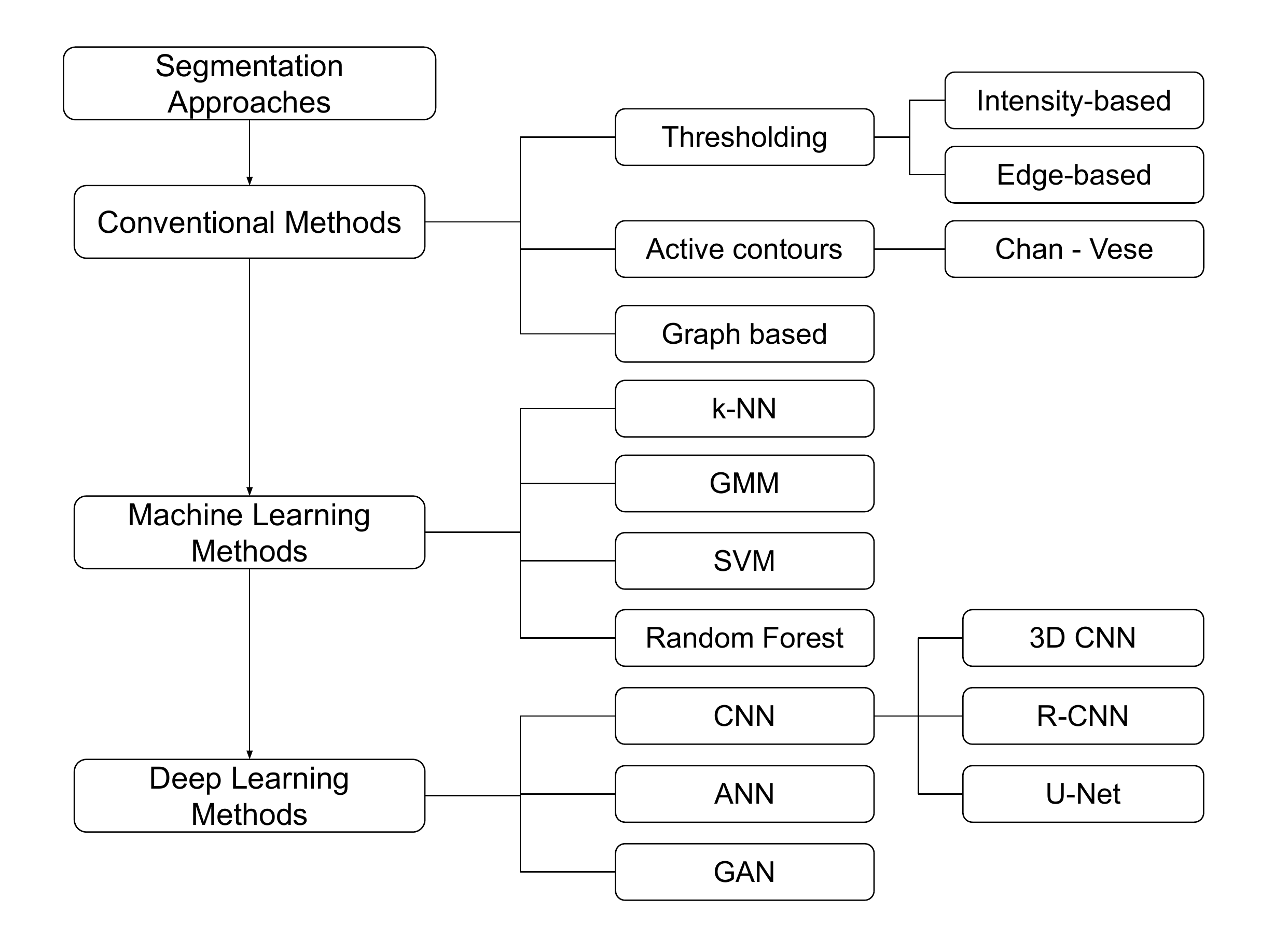}
\caption{Categorization of the reviewed methods.}
\label{fig:diagram}
\end{figure}

All algorithms for automatic segmentation of the \ac{ONH} were analyzed, and the studies were separated in three categories: conventional methods, which use non-learning based image processing techniques only, machine-learning methods (alone or as a refinement/post-process step after conventional methods), and deep learning methods. Figure \ref{fig:diagram} further refines these categories. 

\begin{figure*}
\centering
\includegraphics[width=1\textwidth]{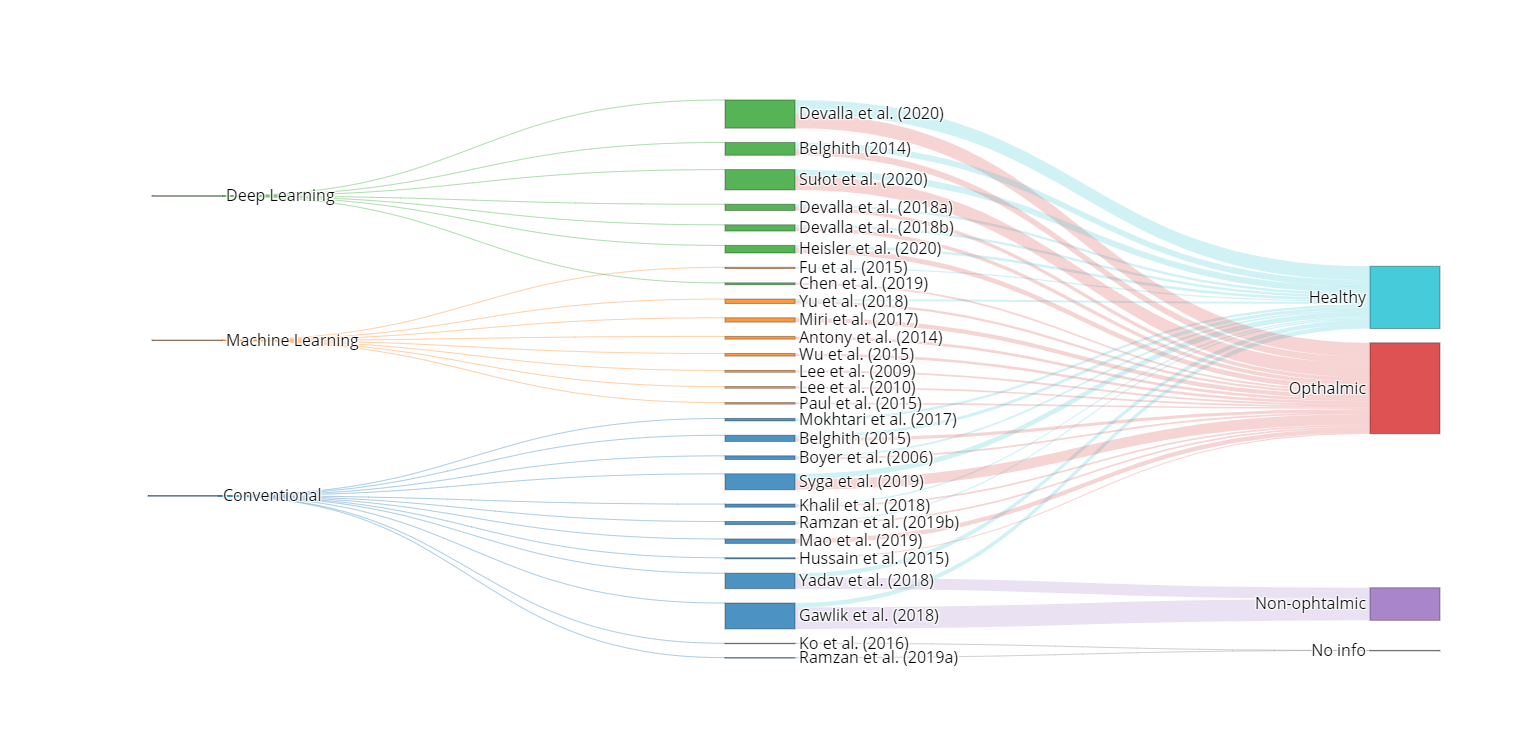}
\caption{Distribution of the reviewed articles organized by method (deep learning, standard machine learning, and conventional image processing) and the type of data used for developing the algorithm (healthy subjects, patients with ophthalmic diseases, and patients with non-ophthalmic diseases). The height of the boxes is proportional the dataset size.}
\label{fig:sankey diagram}
\end{figure*}

The table in the Appendix gives a complete overview of all included studies and their characteristics. The distribution of the papers over the three different categories and the size of the dataset used from ophthalmic and non-ophthalmic patients and healthy subjects in all studies can be found in Figure \ref{fig:sankey diagram}.

Analysing Figure \ref{fig:sankey diagram} it is possible to observe that, overall, deep learning methods use larger datasets.

Most of the reviewed works evaluated the proposed algorithms in pathological data. The only exceptions were \citep{Mokhtari2017}, which validated their method on 40 healthy eyes, and \citep{Ko2016}, which does not specify if the dataset consists of healthy or pathological data. Both authors compare parameters measured in the automatic approaches with manual quantifications. 

Among the rest of the works, glaucoma is the most studied pathology. There are only two works that validate their approaches in pathologies other than glaucoma, \citep{Yadav2018} and \citep{Gawlik2018}, whom applied their method in data from healthy subjects and patients suffering from \ac{IIH}, \ac{MS}, \ac{NMOSD}, and \ac{ON}.

\subsection{Structures of interest}

\begin{figure}[t]
\centering
\includegraphics[width=1\linewidth]{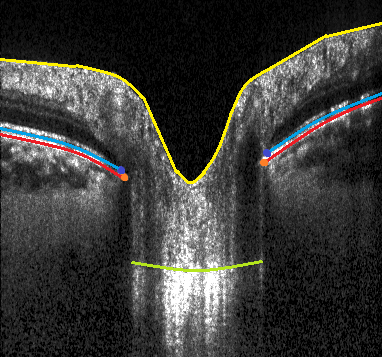}
\caption{\ac{OCT} B-scan (Heidelberg Engineering, Germany) showing a manual boundary-based segmentation of the \ac{ONH}. In yellow, the \ac{ILM} anterior surface; in light blue, the \ac{RPE} layer; in dark blue, the \ac{RPE} endpoints; in red, the Bruch's membrane; in orange, the Bruch's membrane opening points; in green, the \ac{LC} anterior surface.}
\label{fig:manual seg 1}
\end{figure}

\begin{figure}[h]
\centering
\includegraphics[width=\linewidth]{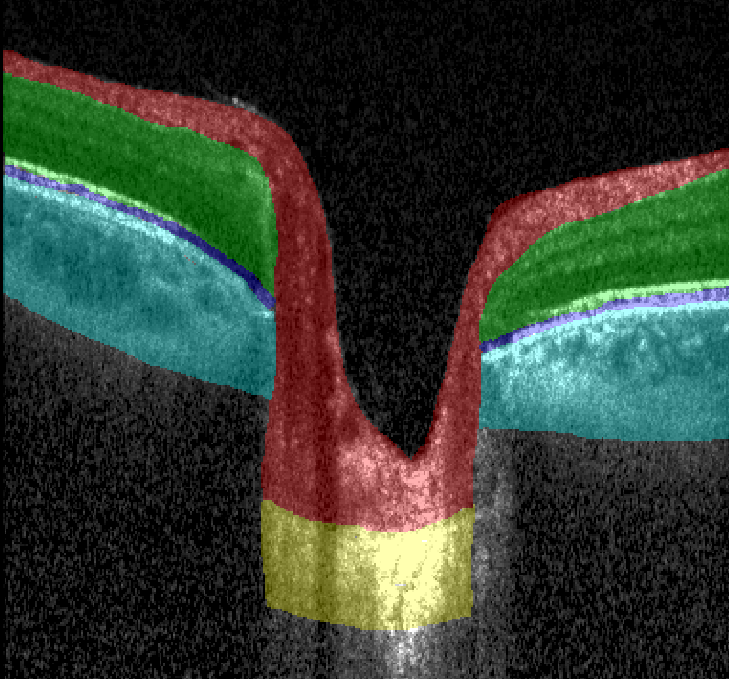}
\caption{\ac{OCT} B-scan (Heidelberg Engineering, Germany) showing a manual region-based segmentation of the \ac{ONH}. In red, the \ac{RNFL}; in green, all retinal layers between the ganglion cell and the photoreceptor layers; in dark blue, the \ac{RPE}; in light blue, the choriocapillaris and choroid; and in  yellow, the \ac{LC}.}
\label{fig:manual seg 2}
\end{figure}

Depending on the aim of the work, different regions or points of interest are segmented in the images. To define these points, two manual segmentations of the \ac{ONH} region in \ac{OCT} images can be found in Figures \ref{fig:manual seg 1} and \ref{fig:manual seg 2} (boundary- and region-based, respectively). Specifically, Figure \ref{fig:manual seg 1} shows the \ac{ILM} anterior surface, the \ac{RPE} layer and its endpoints, the \ac{BM} and its opening points, and the \ac{LC} anterior surface. The vitreal-retinal boundary is equivalent to the anterior surface of the \ac{ILM} layer (shown in yellow).
Figure \ref{fig:manual seg 2} shows the lower boundary of the \ac{RNFL}, and the \ac{LC}. Moreover, it depicts the choroid boundaries. The outer limits of the \ac{ONH} match the retinal-choroidal boundary endpoints. 

In the following subsections, a detailed analysis on the approach used by each algorithm to segment the \ac{ONH} is provided. 

\subsection{Conventional}

Conventional methods are unsupervised segmentation techniques that rely on image processing methods such as thresholding, edge detection and morphological operations. 

\subsubsection{Thresholding}

Thresholding algorithms segment the image based on quantifiable features like image intensity or gradient magnitude, and divide the pixels according to a defined threshold. 

Ramzan \textit{et al.} in \citep{Ramzan2018a} and \citep{Ramzan2018b}, and Khalil \textit{et al.} \citep{Khalil2018} proposed methods to extract the \ac{CDR} through the segmentation of the \ac{ILM} using intensity thresholds. In \citep{Ramzan2018a}, the \ac{ILM} was extracted using intensity based thresholding, followed by interpolation to fill in the gaps. Next, the \ac{RPE} was obtained with Otsu thresholding. Additionally, a novel thresholding approach was introduced. This approach computes the distance to all the centroids of all the objects to remove the extended cup region. In \citep{Ramzan2018b} only the \ac{ILM} was used to calculate the \ac{CDR}. Again, intensity based thresholding was used to detect the \ac{ILM}, defining its anterior surface as the first non-zero layer. Interpolation was applied to fill the gaps of the extracted layer. Finally, Khalil \textit{et al.} \citep{Khalil2018} proposed a method to calculate the \ac{CDR} from the \ac{ILM} segmentation using intensity thresholding, with a refinement that includes interpolation of missing points, outliers removal, low quality image and \ac{RPE} feature analysis. 

Mokhtari \textit{et al.} \citep{Mokhtari2017} used a different approach, applying ridgelet transform \citep{Candes1999} for \ac{RPE} detection and, subsequently, a threshold to determine the \ac{RPE} boundary location.

Boyer \textit{et al.} \citep{Boyer2006} combined thresholding and an adapted Markov model. First, the vitreal-retinal boundary was found through an edge based threshold. A smoothness constraint, that models \textit{a priori} knowledge about the expected outcome of a successful segmentation, was introduced. On most images, below the retinal-vitreal boundary there are two strong dark-to-light edges that can be used to identify outer limits of the \ac{ONH}. These contrasts were used by a Markov model \citep{Koozekanani2001} to extract the full retinal-choroidal boundary.

Also based on edge detection, are the works by Ko \textit{et al.} \citep{Ko2016} and Mao \textit{et al.} \citep{Mao2019}, that used a Canny edge detector as a starting point. In \citep{Ko2016}, Canny edge filter was used to segment the \ac{ONH} surface. After smoothing the surface with a robust version of the local regression, using weighted linear least squares and a second degree polynomial model, each B-scan was re-mapped into a 3D Cartesian domain where the \ac{ONH} and \ac{BMO} surface profiles were fitted with an existing surface modeling tool \citep{Errico2006, DErrico2021}. In \citep{Mao2019}, 2D and 3D Canny edge detectors were applied to the interpolated 3D volume. They used a minimum-cost-path approach, where the cost map was derived from the edge map and the weighted intensity gradients (based on previous knowledge of the \ac{LC} anatomy). The locations with large values of intensity gradient yield the detection the \ac{LC} border. 

\subsubsection{Active contours}

In active contours methods, an object is segmented by an energy-minimizing contour guided by the surrounding pixels. The internal energy comes from the continuity and smoothness of a curve, and the external energy is derived from the edge map of an image \citep{Cheng2020}.

Three of the reviewed works are in this category, two of which \citep{Yadav2018, Gawlik2018} applied Chan-Vese level set approach \citep{Chan2001} as a final step of the method. In \citep{Yadav2018}, active contours were used to upgrade their previous 2D segmentation \citep{Kadas2012a} into a robust 3D segmentation of the \ac{ONH}. They represented the 3D \ac{ONH} shape using triangulated mesh surfaces of the \ac{ILM}, \ac{RPE} and \ac{BMO} points. The identification of the \ac{RPE} lower boundary employs a two-stage thin-plate spline fitting that preserves the retinal natural curvature. Finally, the \ac{ILM} was detected with the Chan-Vese level set approach. The method presented by Gawlik \textit{et al.} \citep{Gawlik2018} presents an extended version of \citep{Yadav2018} by adding a local intensity fitting energy step in order to handle inhomogeneous image intensities.


The third approach based on active contours is the method proposed by Syga \textit{et al.} \citep{Syga2019}, which automatically segmented the \ac{LC} in 3D. Otsu thresholding, morphological operators and interpolation were used to estimate the 3D region of interest in fundus photographs. The obtained information was used to define the elliptic cylindrical region of interest along the z-axis of a cuboid made of B-scans. After \ac{ILM} and \ac{BM} segmentation, in each \ac{OCT} scan, active contours were used to reconstruct the 3D segmentation and find the \ac{LC}.

\subsubsection{Graph based methods}

Graph based methods represent images as a weighted graph, with the pixels of an image as nodes and the relation between the pixels as the arcs or edges of the graph \citep{Singhal}.

Hussain \textit{et al.} \citep{Hussain2015} introduced the approximate location of \ac{TBMR} layers and layers veto (a weight based on the layers pattern) as a parameter of the graph weight function. The goal of their work was to segment the \ac{ONH} and compute the \ac{BMO-MRW} for glaucoma diagnosis. The \ac{TBMR} layers are the three most reflective layers of the retina corresponding to the \ac{RNFL}, the outer neural layer, and the \ac{RPE} and were detected to limit the search space. 

Belghith \textit{et al.} \citep{Belghith2015} proposed a new method to segment the anterior \ac{LC} surface that is able to include prior knowledge in the inference model. They used the \ac{MRF} class of Bayesian methods. 
For segmentation, the \ac{LC} surface was iteratively refined following a perturbation-based approach inspired by the Biased and Filtered Point Sampling method \citep{Chang2011} according to a non-local \ac{MRF} energy function. 

\subsection{Machine Learning}

Machine-learning methods train the algorithm to find patterns and features in large amounts of data in order to make decisions and predictions based on new data. They are usually applied after, or in combination with, conventional methods. 

Several methods in this section use machine learning after a graph search approach. That is the case of the approach followed by Lee \textit{et al.} \citep{Lee2009, Lee2010}, who developed a fast multiscale extension of 3D graph search to detect four intra-retinal surfaces in \ac{ONH} centered \ac{OCT} volumes. The authors used a k-\ac{NN} classifier combined with hull fitting, to segment the \ac{ONH} cup and neuroretinal rim while preserving their shape. The method assigns one of three labels (background, cup, rim) to each column within the \ac{OCT} scan. Then, a set of 15 features was calculated for each voxel column in the volume, and used as input for the k-\ac{NN} classifier. 

Antony \textit{et al.} \citep{Antony2014}, Miri \textit{et al.} \citep{Miri2017}, and Yu \textit{et al.} \citep{Yu2018} all used a random forest classifier \citep{Breiman2001} in combination with the graph search approach. 

In \citep{Antony2014}, an existing graph theoretic approach \citep{Garvin2009} was adapted for simultaneous segmentation of multiple continuous surfaces, in order to make it able to identify the \ac{ONH} boundary in 3D. An iterative method finds the optimal set of feasible surfaces for each estimate of the \ac{ONH} boundary columns by solving a minimum-closure problem in a graph. The random forest classifier was then trained to find the neural canal opening boundary points, based on the previously learned textural features. However, the continuation of the iterative search on the border of the \ac{BM} was a common mistake that lead to a wrong placement of the \ac{BMO} and, consequently, of the borders of the \ac{ONH}. This limitation was addressed in \citep{Miri2017} by eliminating the iteration phase. The new method, instead of using a mathematical model for the \ac{BMO} points, computes the likelihood of each voxel being a \ac{BMO} point using the random forest classifier. 

The method described in \citep{Yu2018} includes locally adaptive constraints for a more accurate \ac{ONH} region detection.
The \ac{ONH} region was first detected by random forest method on polar-transformed images with features representing both textural and structural information. Then two layer segmentation methods with locally adaptive constraints, Otsu segmentation guided graph search and shared-hole graph search, were proposed for the segmentation of nine surfaces.

The three remaining papers of this section proposed all different approaches. Fu \textit{et al.} \citep{Fu2015} detects the \ac{OD} through the segmentation of the \ac{RPE}. A low-rank dictionary based on intensity features and local binary patterns was learned and used to reconstruct the layer on the candidate region. The resulting error curves, that represent the deviations from the smooth geometrical structure, allowed for the boundaries of the \ac{OD} to be detected. Paul \textit{et al.} \citep{Paul2015} performed a segmentation of the retinal and vitreal boundary from \ac{OCT} \ac{ONH} centered images by incorporating a \ac{GMM} \citep{reynolds2009gaussian} clustering into a kernel. Finally, Wu \textit{et al.} \citep{Wu2015} started by using a multi-scale 3D graph search approach to segment the \ac{RPE}, followed by a search patch method to segment the \ac{ONH}. A \ac{SVM} classifier was trained with the purpose of finding the most likely patch centered at the neural canal opening. The features extracted for patch description were the local binary pattern and histogram of gradient.

\subsection{Deep Learning}

Deep learning methods are an advanced type of machine-learning algorithms that have been gaining visibility in the last decade. Following computational developments, they are capable of extracting and classifying features automatically when a large amount of training data is given \citep{Seo}. The most commonly used architectures for medical image segmentation are based on the U-net \citep{Ronneberger}.

Belghith \textit{et al.} \citep{Belghith2014} addressed the segmentation problem by improving an existing machine learning based method by Lee \textit{et al.} \citep{Lee2010}. In  \citep{Lee2010}, the estimation of the layers highly depends on the accuracy of the estimated gradient-based transitions, which can be a major drawback for low quality and noisy images, particularly in the \ac{BMO} area. To overcome this, the authors proposed the use of a \ac{ANN} and \ac{PCA} \citep{kirby1996circular}. That way, the elliptical shape of the \ac{BM} curve can be modeled, obtaining a more accurate estimate of the \ac{ONH} size. 

Among the authors that used a U-net as a basis for their algorithm are Chen \textit{et al.} \citep{Chen2019}, who proposed a method consisting of three steps. First, a coarse detection based on the \ac{RPE} layer and \ac{ONH} segmentation in 2D projection image was applied. Then, a U-net was used to improve the accuracy of the coarse detection. Finally, a post processing algorithm removes the outliers. The loss function was a combination of Dice loss with an area bias and the mean square error loss. 

In \citep{Skander2020}, the aim was to segment the \ac{BMO}. To that end, three deep learning based approaches were used and compared while evaluating the effect of the input size: an \ac{ANN} where the input is an A-scan, a patch based \ac{CNN} method where the input is a group of consecutive A-scans and a U-net where the input is a B-scan.

In \citep{Heisler2020}, the aim was layer segmentation. The proposed method, a semi-supervised \ac{GAN} \citep{Isola2017}, allowed for the training of the network with smaller datasets while taking advantage of unlabeled scans. Additionally, a Faster Region \ac{CNN} \citep{Ren2016} was used to segment the \ac{BMO} from the volumes.

The last three deep learning approaches, all developed by Devalla \textit{et al.} \citep{Devalla2017, Devalla2018, Devalla2020}, present different architectures for segmentation. The aim in \citep{Devalla2017} was to segment six neural and connective tissue structures in \ac{OCT} images of the \ac{ONH}: the \ac{RNFL} and the prelamina, the \ac{RPE}, the set of remaining retinal layers, the choroid, the peripapillary sclera, and the \ac{LC}. After pre-processing all B-scans with adaptive compensation \citep{Mari2013a}, they used a two dimensional \ac{CNN} that was trained with manually segmented images. This approach does not offer a precise separation between the \ac{LC} and the sclera. This drawback was tackled in \citep{Devalla2018}, which proposes an architecture combining a U-net and residual blocks. The goal is to extract the same six regions  of the \ac{ONH} by capturing both the contextual and local information while taking advantage of residual connections to improve the flow of the gradient information through the network. When compared to \citep{Devalla2017}, the results showed that the new architecture performed better for all the tissues except for the\ac{RPE}, where they performed similarly.

Finally, the work in \citep{Devalla2020} attempted to address one of the main obstacles for the automatic segmentation of the \ac{ONH} in the clinical environment: the lack of device-independent segmentation algorithms. One of the key elements of the proposed framework was a pre-processing enhancement step, which makes use of a deep learning network to improve the quality of \ac{OCT} B-scans and to harmonize image characteristics across \ac{OCT} devices.

The authors \citep{Devalla2020} found that the use of 3D \ac{CNN}s could further improve the reliability of the automatic segmentation by considering depth-wise spatial information from adjacent images. The proposed architecture combined three segmentation \ac{CNN}s based on the 3D U-net. Each of the three 3D \ac{CNN}s offered an equally plausible segmentation. However, the segmentation of ambiguous regions, such as the sclera and the \ac{LC}, can differ considerably between networks with different structures. Therefore, an ensembler was used to combine the predictions from the three networks, giving a more robust segmentation in the end.

\subsection{Evaluation and validation}

\input{boundary_based}

A summary of the most representative results of the reviewed works can be found in Tables \ref{tab:boundary based} and \ref{tab:region based}. Several outcomes can be used to evaluate a segmentation. These outcomes can be either a segmented tissue, such as the \ac{ILM}, the \ac{LC}, the \ac{RNFL}, the \ac{RPE}, the other retinal layers and the choroid, or it can be a biomarker related to a segmented tissue, such as the \ac{BMO} and optic cup detection, the \ac{BMO-MRW}, the \ac{CDR} and the \ac{ONH} surface depth. For each of these outcomes the metrics used for their evaluation and the quantitative results reported are presented.

One of the metrics mentioned in table \ref{tab:boundary based} is the failure rate. This metric was proposed by Belghith \textit{et al.} \citep{Belghith2014, Belghith2015} and it compares the automatic segmentation with the ground truth. A failure rate of 0 is obtained when the mean difference $<$3 pixels, of 1 when the mean difference $<$5 pixels and of 2 when the mean difference $>$5 pixels.

\input{region_based}

Even though three works reported a region segmentation of the \ac{LC} \citep{Devalla2018, Devalla2018a, Devalla2020}, no values are reported in Table \ref{tab:region based}. Given the subjectivity in the visibility of the posterior \ac{LC} boundary, the groups were only able to do a qualitative assessment of this the segmentation of the \ac{LC}.

From the analysed pool, two of the works studied not only if the qualify of the segmentation was good, but also if the parameters obtained from the automatic segmentation were able to correctly classify data from the different groups. Ramzan  \textit{et al.} \citep{Ramzan2018a} evaluated the performance of the computed \ac{CDR} in separating healthy from glaucomatous eyes, and it showed an average sensitivity of 87\%, specificity of 73\%, and accuracy of 79\%. Syga \textit{et al.} \citep{Syga2019} validated their model on a dataset from 255 subjects, obtained a 68\% accuracy and 0.66 \ac{AUC} in distinguishing \ac{POAG} patients from controls (p-value $<$0.001), 64\% and 0.585 between suspects with \ac{GODA} and controls (p-value $<$0.015), and 56\% and 0.561 between patients and suspects (p-value $=$ 0.333) based on the mean \ac{LC} index (total shape of the \ac{LC} parameterization based on the fourth-order polynomial fit). 

\section{Discussion}

The present review collects and summarizes the existing automatic algorithms for the segmentation of the \ac{ONH} in \ac{OCT} scans. It shows that improvements are necessary in the field since there is a limited number of studies, with great diversity in the size and type of datasets used, segmented regions and validation methods, which precludes a comparison between studies. Boundary segmentation was the starting point for the detection of the \ac{ONH} and its layers. However, as methods developed, region segmentation has also been proposed.

Conventional methods focused mainly on segmenting boundaries, through the detection of \ac{BMO} points, the \ac{ILM} and the \ac{RPE} limits. For the \ac{LC}, only the anterior surface could be detected, which limited the parameters that could be extracted with these methods.

In medical imaging segmentation tasks, it is often assumed that the surfaces are continuous, which is not the case for the \ac{ONH} in which surfaces converge to a hole. This can be a problem when segmenting structures with multiple interacting surfaces, such as in \ac{OCT} volumes of the \ac{ONH}.

Since machine-learning methods were mostly applied after conventional methods, they were often able to address segmentation problems that had remained unanswered with prior methods. Particularly, for accurately identifying the optic cup, for which pattern recognition played an important role.  Miri \textit{et al.} \citep{Miri2017} were able to improve the unsigned border error of \ac{BMO-MRW} from previous methods by at least 4 $\mu$m (26.65 $\pm$ 13.27 $\mu$m and 22.22 $\pm$ 5.99 $\mu$m). 

Deep learning methods have been gaining visibility for their success in other medical imaging processing and analysis, and can be the future of research in this field \citep{RizwanIHaque2020}. 
Using deep learning, \citep{Chen2019} was able to outperform the mean error of previous \ac{BMO} segmentation by at least 7 $\mu$m.
Moreover, only deep learning based algorithms have been able to perform a region segmentation of several tissues of the \ac{ONH}.
Devalla \textit{et al.} were able to accurately segment almost all connective and neural tissues with sensitivities and specificities around or above 90\%, except for the \ac{LC}, that despite improvements, remains a challenge due to low signal-to-noise ratio.

Connective tissues, such as the peripapillary sclera, \ac{BM} and the \ac{LC} are the main load bearing elements of the \ac{ONH}. Parameters extracted from the segmentation of these type of tissues have already proven to make a difference in several diagnoses \citep{Yang2015}. The \ac{LC} load-bearing connective tissue components comprise about 40\% of the tissue volume in the laminar region of the \ac{ONH} \citep{Downs2017}.
Adding to its anatomical location, the \ac{LC} becomes a weak spot with the conflicting tasks of providing structural and nutrient support to the axons while withstanding mechanical strain \citep{Downs2017}. When compressed above a certain point, the \ac{LC} can be deformed, compromising axonal transport and tissue remodeling by reactive astrocytes \citep{Lee2017}, as well as the diffusion of nutrients from the capillaries \citep{Burgoyne2005}.
Several studies have already shown that \ac{LC} features, such as \ac{LC} depth and thickness, have potential to be used in clinical diagnosis \citep{Paulo2021}. Moreover, being significantly different between healthy patients and ocular and systemic pathologies while being patient-specific features \citep{Paulo2021}, \ac{LC} features are seen as increasingly promising for patient follow-up as well. Therefore, the remaining lack of accuracy in detecting the \ac{LC} can affect diagnosis and follow-up.

Altogether, the values of the parameters extracted from the segmentations showed significant differences between healthy and pathological groups. Methods such as the ones developed by Ramzan \textit{et al.} \citep{Ramzan2018a} and Syga \textit{et al.} \citep{Syga2019} already achieved sensitivities of 87\% and 81\%  in distinguishing healthy from pathological groups. 

Datasets with data from less than 40 patients were often used. Even though a lot of imaging data are being acquired in clinical practice, these data are rarely labelled and/or publicly available. The time consuming process of manual labelling by experts, combined with the scarcity of publicly available segmented datasets, may cause further delays in technology development, since time is lost in repeating steps that have already been done and validated by previous groups. 
This is particularly problematic for supervised deep learning methods since they need more data to yield accurate results. However, future work may focus on label-free/unsupervised learning since it will ease the burden of manual labelling (however, it will not solve the lack of \ac{OCT} data itself). Therefore, efforts to make more clinical data available and create sharing practices/protocols between groups could further accelerate research, allowing more studies to be done more effectively, which will close the gap to automation in the clinic. 

One limitation of this review is that, since the literature search was made on MEDLINE (Pubmed) and Scopus bibliographic databases only, some technical studies might have been missed. By considering only articles with detailed descriptions on the algorithm used, the number of included articles was shortened since otherwise a review of the method would not be possible. Moreover, studies which were solely published as congress abstracts were excluded from this review.

\section{Conclusion}
There is a growing interest in \ac{ONH} features as biomarkers for disease diagnosis and/or progression. This review highlights algorithms that automatically segment several structures and boundaries from \ac{ONH} centered \ac{OCT} scans. From these automatic segmentations, several parameters can be automatically extracted which may be relevant for clinical practice. Nevertheless, efforts should be employed to make more \ac{OCT} data available, develop standardized guidelines for the extracted parameters and metrics used in the validation of the algorithms so that more accurate comparisons between methods can be performed. Moreover, efforts in improving \ac{LC} signal-to-noise ratio and device-independent algorithms can contribute to a better diagnosis and follow-up of ONH-related diseases in daily clinical practice.

\bibliographystyle{plainnat}
\bibliography{mainRita2}{}

\newpage
\onecolumn
\global\pdfpageattr\expandafter{\the\pdfpageattr/Rotate 90} 
\input{Summary_Table}

\end{document}

%% file: boundary_based.tex
\begin{table*}[h]
\centering
\caption{Results from the boundary based segmentations.}
\label{tab:boundary based}
\resizebox{\textwidth}{!}{%
\begin{tabular}{lll}
\hline
\multicolumn{1}{c}{Outcome} &
  \multicolumn{1}{c}{Metrics} &
  \multicolumn{1}{c}{Results} \\ \hline
BMO &
  \begin{tabular}[c]{@{}l@{}}(i) Correlation with ground truth; \\ \\ (ii) Error compared with ground truth; \\ \\ (iii) Mean unsigned error; \\ \\ (iv) Mean signed error;\\ \\ (v) Dice similarity coefficient;\\ \\ (vi) Failure rate;\\ \\ (vii) Mean average precision\end{tabular} &
  \begin{tabular}[c]{@{}l@{}}(i) 0.93 \citep{Boyer2006}; \\ (ii) 32.03 $\pm$ 58.68 $\mu$m \citep{Mokhtari2017}. 12.4 $\pm$ 12.1 pixels \citep{Fu2015}. \\ 2.8 pixels (normal) and 3.1 pixels (glaucoma) \citep{Paul2015}. \\ 54.18 $\pm$ 53.74 $\mu$m \citep{Hussain2015}. \\ 60.00 $\pm$ 42.00 $\mu$m \citep{Wu2015}. 49.28 $\pm$ 16.78 $\mu$m \citep{Miri2017}. \\ 42.38 $\pm$ 18.33 $\mu$m \citep{Chen2019}.\\ (iii) 61.86 $\pm$ 61.97 $\mu$m (x axis) and 12.40 $\pm$ 11.24 $\mu$m (z axis) \citep{Yadav2018}. \\ 49.53 $\pm$ 30.41 $\mu$m (x axis) and 31.58 $\pm$ 21.06 $\mu$m (z axis) \citep{Antony2014}. \\ 37.98 $\pm$ 14.91 $\mu$m (x axis) and 22.28 $\pm$ 8.58 $\mu$m (z axis) \citep{Miri2017}. \\ 9.80 $\pm$ 31.90 $\mu$m \citep{Skander2020}. \\ 0.023 mm \citep{Lee2009}. 0.026 mm \citep{Lee2010}\\ (iv)  -7.69 $\pm$ 87.27 $\mu$m (x axis) and -1.41 $\pm$ 16.69 $\mu$m (z axis) \citep{Yadav2018}. \\ 26.49 $\pm$ 40.22 $\mu$m (x axis) and 25.45 $m$ 14.37 $\mu$m (z axis) \citep{Antony2014}.\\  -9.49 $\pm$ 24.58 $\mu$m (x axis) and 8.33 $\pm$ 17.72 $\mu$m (z axis) \citep{Miri2017}.\\ (v) 0.925 $\pm$ 0.030 \citep{Yu2018}. 0.959 $\pm$ 0.032 \citep{Skander2020}. \\ 0.65 $\pm$ 0.14 \citep{Lee2010}.\\ (vi) 0 in 92.5\% of the scans \citep{Belghith2014}\\ (vii) 0.8547 for glaucoma and 0.9567 for control subjects \citep{Heisler2020}.\end{tabular} \\ \hline
BMO-MRW &
  \begin{tabular}[c]{@{}l@{}}(i) Mean unsigned error;\\ (ii) Mean signed error; \\ (iii) RMSE\end{tabular} &
  \begin{tabular}[c]{@{}l@{}}(i) 58.62 $\pm$ 43.12 $\mu$m \citep{Hussain2015}. 26.65 $\pm$ 13.27 $\mu$m \citep{Antony2014}.\\  22.22 $\pm$ 5.99 $\mu$m \citep{Miri2017}.\\ (ii) 6.61 $\pm$ 18.59 $\mu$m \citep{Antony2014}. -0.30 $\pm$ 12.44 $\mu$m \citep{Miri2017}.\\ (iii) 17.99 $\pm$ 8.15 $\mu$m \citep{Antony2014}. 11.62 $\pm$ 4.63 $\mu$m \citep{Miri2017}\end{tabular} \\ \hline
Optic Cup &
  \begin{tabular}[c]{@{}l@{}}(i) Correlation with ground truth;   \\ (ii) Error compared with ground truth;\\ (iii) Mean unsigned error;\\ (iv) Dice similarity coefficient\end{tabular} &
  \begin{tabular}[c]{@{}l@{}}(i) 0.80 \citep{Boyer2006}\\ (ii) 3.4 pixels (normal) and 3.6 pixels (glaucoma) \citep{Paul2015}\\ (iii) 0.009 mm \citep{Lee2009}. 0.038 mm \citep{Lee2010}\\ (iv) 0.85 $\pm$ 0.06 \citep{Lee2010}\end{tabular} \\ \hline
CDR &
  \begin{tabular}[c]{@{}l@{}}(i) Error compared with ground truth;\\ (ii) Sensitivity;\\ (iii) Specificity;\\ (iv) Accuracy\end{tabular} &
  \begin{tabular}[c]{@{}l@{}}(i) 0.045 $\pm$ 0.033 \citep{Wu2015} \\ (ii) 86.2 $\pm$ 7.1\% \citep{Khalil2018}\\ (iii) 86.2 $\pm$ 9.8\% \citep{Khalil2018}\\ (iv) 85.5 $\pm$ 5.2\% \citep{Khalil2018}\end{tabular} \\ \hline
ILM &
  \begin{tabular}[c]{@{}l@{}}(i) Mean unsigned error; \\ (ii) Mean Dice coefficient;\\ (iii) Error compared with ground truth;\\ (iv) Mean Euclidean distance error\end{tabular} &
  \begin{tabular}[c]{@{}l@{}}(i) 5.38 $\pm$ 4.23 $\mu$m \citep{Yu2018}.\\ (ii) 0.97 \citep{Heisler2020}.\\ (iii) 4.80 $\mu$m \citep{Gawlik2018}\\ (iv) 7.08 $\pm$ 3.7 pixels \citep{Ramzan2018a}. 7.03 $\pm$ 3.72 pixels\end{tabular} \\ \hline
LC &
  \begin{tabular}[c]{@{}l@{}}(i) Failure rate;  \\ (ii) Accuracy\end{tabular} &
  \begin{tabular}[c]{@{}l@{}}(i) 0 in 73.7\% of the scans \citep{Belghith2015} \\ (ii) 90.6\% \citep{Mao2019}\end{tabular} \\ \hline
ONH surface depth &
  (i) Error compared with ground truth &
  (i)  0.7 $\pm$ 1.0 \% \citep{Ko2016}. \\ \hline
\end{tabular}%
}

\end{table*}

%% file: region_based.tex
\begin{table*}[h]
\centering
\caption{Results from the region based segmentation}
\label{tab:region based}
\resizebox{\textwidth}{!}{%
\begin{tabular}{lllc}
\hline
\multicolumn{1}{c}{Outcome} &
  \multicolumn{1}{c}{Metrics} &
  \multicolumn{2}{c}{Results} \\ \hline
RNFL + prelamina &
  \begin{tabular}[c]{@{}l@{}}(i) Dice coefficient;\\ (ii) Sensitivity;\\ (iii) Specificity;\\ (iv) Accuracy\end{tabular} &
  \begin{tabular}[c]{@{}l@{}}(i) 0.82 $\pm$ 0.05 \citep{Devalla2018a}. 0.92 $\pm$ 0.05 for healthy \\ and 0.92 $\pm$ 0.03 for glaucoma \citep{Devalla2018}. \\ (ii) 0.89 $\pm$ 0.04 \citep{Devalla2018a}\\ (iii) 0.99 \citep{Devalla2018a}. 0.99 \citep{Devalla2018}.\\ (iv) 0.93 $\pm$ 0.02 \citep{Devalla2018a}\end{tabular} &
  \multirow{15}{*}{\begin{tabular}[c]{@{}c@{}}  Mean for all tissues:\\ (ii) 0.94 $\pm$ 0.02 (Spectralis), \\ 0.93 $\pm$ 0.02 (Cirrus) and \\0.93 $\pm$ 0.02 (RTVue) \citep{Devalla2020}.\\ (iii) 0.99 for Spectralis, Cirrus \\ and RTVue \citep{Devalla2020}.\end{tabular}} \\ \cline{1-3}
RPE &
  \begin{tabular}[c]{@{}l@{}}(i) Dice coefficient \\ (ii) Sensitivity;\\ (iii) Specificity;\\ (iv) Accuracy\end{tabular} &
  \begin{tabular}[c]{@{}l@{}}(i) 0.84 $\pm$ 0.02  \citep{Devalla2018a}. 0.83 $\pm$ 0.04 for healthy  \\ and 0.84 $\pm$ 0.03 for glaucoma \citep{Devalla2018}.\\ (ii) 0.90 $\pm$ 0.03 \citep{Devalla2018a}\\ (iii) 0.99 \citep{Devalla2018a}. 0.99 \citep{Devalla2018}.\\ (iv) 0.93 $\pm$ 0.02 \citep{Devalla2018a}\end{tabular} &
   \\ \cline{1-3}
Other retinal layers &
  \begin{tabular}[c]{@{}l@{}}(i) Dice coefficient\\ (ii) Sensitivity;\\ (iii) Specificity;\\ (iv) Accuracy\end{tabular} &
  \begin{tabular}[c]{@{}l@{}}(i) 0.86 $\pm$ 0.03  \citep{Devalla2018a}. 0.95 $\pm$ 0.01 for healthy \\ and 0.96 $\pm$ 0.03 for glaucoma \citep{Devalla2018}.\\ (ii) 0.98 $\pm$ 0.02 \citep{Devalla2018a}\\ (iii) 0.99 \citep{Devalla2018a}. 0.99 \citep{Devalla2018}.\\ (iv) 0.98 $\pm$ 0.01 \citep{Devalla2018a}\end{tabular} &
   \\ \cline{1-3}
Choroid &
  \begin{tabular}[c]{@{}l@{}}(i) Dice coefficient\\ (ii) Sensitivity;\\ (iii) Specificity;\\ (iv) Accuracy\end{tabular} &
  \begin{tabular}[c]{@{}l@{}}(i) 0.85 $\pm$ 0.02 \citep{Devalla2018a}. 0.90 $\pm$ 0.03 for healthy \\ and 0.91 $\pm$ 0.05 for glaucoma \citep{Devalla2018}.\\ (ii) 0.91 $\pm$ 0.02 \citep{Devalla2018a}\\ (iii) 0.99 \citep{Devalla2018a}. 0.99 \citep{Devalla2018}.\\ (iv) 0.93 $\pm$ 0.01 \citep{Devalla2018a}\end{tabular} &
   \\ \hline
\end{tabular}%
}

\end{table*}

%% file: Summary_Table.tex
\def\arraystretch{1.5}
\newcolumntype{C}[1]{>{\centering\arraybackslash}m{#1}}
\begin{scriptsize}
\begin{landscape}
\begin{longtable}{C{1.8cm}C{1.5cm}C{3cm}C{3.5cm}C{4.5cm}C{4.5cm}C{1cm}C{2.5cm}}
\caption{Characteristics of the reviewed studies}\\ \hline
\textbf{Category} &
  \textbf{Authors} &
  \textbf{Dataset} &
  \textbf{Regions Segmented} &
  \textbf{Validation} &
  \textbf{Results} &
  \textbf{Technique} &
  \textbf{Device} \\ \hline
\endhead

  Conventional &
  \citet{Boyer2006} &
  59 glaucoma B-scans  &
  Retinal-vitreal boundary, limits of the \ac{OD}, retinal-choroid boundaries. &
  Comparison with the ground truth. &
  High correlation between automatic and manual cup and disk limits. &
  - &
  OCT 3000 from Zeiss-Humphrey \\ \hline 

  Conventional &
  \citet{Hussain2015} &
 13 glaucoma scans &
  ILM, BMO, HRC, RNFL, RPE, \ac{OD} boundary &
  Comparison with ground truth and existing method. Confusion matrix based metrics. Distance metrics.&
  Robust segmentation over noise and pathology.&
  SD-OCT &
  Spectralis \\ \hline  
 
  Conventional &
  \citet{Belghith2015} &
  50 healthy scans and 50 glaucoma scans &
  LC anterior surface &
  Comparison with ground truth.  Statistical tests. &
  High similarity between manual and automatic segmentation. Significant correlation between changes in IOP and the position of the LC. &
  EDI SD-OCT &
  Spectralis \\ \hline   

  Conventional &
  \citet{Ko2016} &
  no info &
  ILM, BMO &
  Comparison with ground truth.  &
  Accurate segmentation of the ONH structure. Not sensitive to the differentiation of blood vessels from the ONH surface. &
  SD-OCT &
  Spectralis \\ \hline  

 Conventional &
  \citet{Mokhtari2017} &
  40 healthy scans &
  RPE break points, RPE boundary &
  Comparison with ground truth. &
  Accurate segmentation of the \ac{OD} boundary. &
  no info &
  Topcon model of 3D-1000 unit \\ \hline   

Conventional &
  \citet{Gawlik2018} &
  71 healthy scans and 345 pathological scans (31 IIH + 60 NMOSD + 252 MS) &
  ILM &
  Statistical tests. Comparison with segmentation from a device. Distance metrics. Visual evaluation.&
  Robust segmentations over variations in ONH topology. Outperforms device segmentation.&
  SD-OCT &
  Spectralis \\ \hline  

Conventional &
  \citet{Khalil2018} &
  22 healthy scans and 28 glaucoma scans  &
  ILM and RPE &
  Comparison with ground truth. Comparison with computed generated values. Comparison with existing methods \citet{wang2013quantitative, nithya2015analysis, babu2015optic}.&
  Outperforms existing methods and computer generated values.&
  SD-OCT &
  no info \\ \hline  
 
 Conventional &
  \citet{Yadav2018} &
  71 healthy scans and 177  pathological scans (31 IIH +146 autoimmune central nervous system disorders ) &
  RPE, BMO Points, ILM and BMO. &
  Comparison with ground truth. Distance metrics. &
  Successfully captures the differences between pathological groups.&
  SD-OCT &
  Spectralis \\ \hline
  
  Conventional &
  \citet{Ramzan2018a} &
  50 scans from healthy and glaucoma patients &
  ILM and RPE &
  Comparison with ground truth. comparison with computer generated values. Confusion matrix based metrics.&
  High correlation with ground truth. Outperforms existing techniques.&
  TOPCON'S 3D OCT-1000 \\ \hline 
 
 Conventional &
  \citet{Ramzan2018b} &
  50 scans from healthy and glaucoma patients &
  ILM &
  Comparison with ground truth. Distance metrics. Visual evaluation. &
  Accurate segmentation. &
  SD-OCT &
  Topcon \\ \hline 
 
 Conventional &
  \citet{Syga2019} &
  86 healthy scans and 169 glaucoma scans &
  ILM, BM points and LC &
  Statistical tests. Confusion matrix based metrics. &
  Statiscally significant differences between glaucoma patients (POAG and GODA) and healthy controls. The mean LC shapes for POAG and GODA were not significantly different.&
  EDI OCT &
  Spectralis \\ \hline 
  
Conventional &
  \citet{Mao2019} &
  72 glaucoma scans &
  LC anterior surface &
  Comparison with ground truth. Visual evaluation.&
  Segmentation accuracy is significantly higher when a deep learning noise reduction algorithm is used than in raw images.&
  SS-OCT &
  Topcon \\ \hline
 
 Machine learning &
  \citet{Lee2009} &
  30 glaucoma scans  &
  3 intraretinal surfaces, \ac{OD} boundary &
  Comparison with ground truth. Distance metrics. &
  No signifficant differences between the unsigned errors of the optic cup and disk, before and after feature selection. &
  HD-OCT &
  Cirrus \\ \hline
  
  Machine learning &
  \citet{Lee2010} &
  27 glaucoma scans &
  4 intraretinal surfaces, \ac{OD} boundary &
  Comparison with the ground truth. Distance metrics. Confusion matrix based metrics. &
  Contextual 9-k-NN outperforms the regular k-NN classifier when no post processing is applied. Performance of 9-k-NN classifier is significantly better with post processing.&
  HD-OCT &
  Cirrus \\ \hline 
 
  Machine learning &
  \citet{Antony2014} &
  44 glaucoma scans  &
  BM, intraretinal surfaces, ONH hole &
  Comparison with ground truth and existing methods. Distance metrics. Confusion matrix based metrics.&
  Outperforms existing methods \citet{hu2010automated, Lee2010}. &
  SD-OCT &
  Cirrus \\ \hline
 
  Machine learning &
  \citet{Wu2015} &
  42 glaucoma scans  &
  ILM, RPE &
  Comparison with ground truth and existing method. Distance metrics. Confusion matrix based metrics. &
  Outperforms existing methods \citet{Lee2010}. &
  SD-OCT &
  Cirrus \\ \hline
 
  Machine learning &
  \citet{Fu2015} &
  48 healthy scans  &
  ILM, RPE, \ac{OD} boundary &
  Comparison with ground truth and other methods.  &
  Outperforms existing methods \citet{Boyer2006}. &
  SD-OCT &
  Topcon \\ \hline 
 
  Machine learning &
  \citet{Paul2015} &
  25 glaucoma B-scans &
  Retinal layers, \ac{OD} boundary &
  Comparison with ground truth. &
  Correct segmentation in normal and glaucoma affected images.&
  no info &
  no info \\ \hline
 
  Machine learning &
  \citet{Miri2017} &
  69 glaucoma scans &
  ILM, (IS/OS), BM &
  Comparison with ground truth and existing methods. Distance metrics. Statistical tests. &
  Outperforms existing methods \citet{Antony2014}. &
  HD-OCT &
  Cirrus \\ \hline

   Machine learning &
  \citet{Yu2018} &
  30 healthy scans and 35 glaucoma scans &
  RNFL, retinal layers, RPE/Bruch’s complex and \ac{OD} boundary &
  Comparison with ground truth and existing methods. Confusion matrix based metrics. Distance metrics.&
  Outperforms existing methods \cite{Hu2009, zang2017automated}. &
  SD-OCT &
  Topcon \\ \hline
 
 Deep learning &
  \citet{Belghith2014} &
  100 healthy scans and 105 glaucoma scans &
  BM &
  Comparison with ground truth, values from the device and other methods. Statistical tests.&
  High correlation with ground truth and built-in software of the device. Significant differences between glaucoma and healthy eyes.&
  EDI SD-OCT &
  Cirrus and Spectralis \\ \hline 
 
 Deep learning &
  \citet{Devalla2018a} &
  40 healthy scans and 60 glaucoma scans &
  RNFL and the prelamina; RPE; all other retinal layers; the choroid; the peripapillary sclera and the LC; &
  Confusion matrix based metrics. Statistical tests.&
  Good performance for all tissues in glaucoma and healthy images. Performs better with compensated images.&
  EDI SD-OCT &
  Spectralis \\ \hline
 
 Deep learning &
  \citet{Devalla2018} &
  40 healthy scans and 60 glaucoma scans &
  RNFL and the prelamina; RPE; all other retinal layers; choroid; peripapillary sclera; LC; &
  Confusion matrix based metrics. &
  Good performance for all tissues in glaucoma and healthy images. No significant differences in segmentation performances with compensated and uncompensated images.&
  EDI OCT &
  Spectralis \\ \hline
 
 Deep learning &
  \citet{Chen2019} &
  30 glaucoma scans &
  \ac{OD} boundary, RPE, BMO points &
  Comparison with the ground truth and with existing methods.&
  Outperforms existing methods \cite{Hussain2015, Wu2015, Miri2017}. &
  SD-OCT &
  Topcon \\ \hline
 
 Deep learning &
  \citet{Heisler2020} &
  42 healthy scans and 80 glaucoma scans &
  ILM, RNFL, BM, choroid-sclera boundary; BMO points &
  Comparison with ground truth. Confusion matrix based metrics. Statistical tests. &
  No statistically significant difference between BMO segmentation and ground truth. Thickness parameters were highly correlated.&
  SS-OCT &
  custom-built OCT \\ \hline
 
 Deep learning &
  \citet{Devalla2020} &
  225 healthy scans and 225 glaucoma scans &
  RNFL and prelamina; ganglion cell complex; all other retinal layers; RPE; choroid; LC &
  Confusion matrix based metrics. Visual evaluation. Statistical tests. &
  Networks trained in any of the devices, successfully segmented images from other devices with high performances in all tissues.&
  SD-OCT &
  Spectralis, Cirrus and RTVue \\ \hline

  Deep learning &
\citet{Skander2020} &
  102 healthy scans and 223 glaucoma scans &
  BMO &
  Comparison with ground truth. Confusion matrix based metrics. Distance metrics.&
  U-net like architecture with B-scans as input had the best performance.&
  SD-OCT &
  Spectralis \\ \hline

\end{longtable}
\end{landscape}
\end{scriptsize}